# Thermal hyperbolic metamaterials

**Yu Guo and Zubin Jacob***
Department of Electrical and Computer Engineering
University of Alberta, Edmonton, AB T6G 2V4, Canada
*zjacob@ualberta.ca*

**Abstract:** We explore the near-field radiative thermal energy transfer properties of hyperbolic metamaterials. The presence of unique electromagnetic states in a broad bandwidth leads to super-planckian thermal energy transfer between metamaterials separated by a nano-gap. We consider practical phonon-polaritonic metamaterials for thermal engineering in the mid-infrared range and show that the effect exists in spite of the losses, absorption and finite unit cell size. For thermophotovoltaic energy conversion applications requiring energy transfer in the near-infrared range we introduce high temperature hyperbolic metamaterials based on plasmonic materials with a high melting point. Our work paves the way for practical high temperature radiative thermal energy transfer applications of hyperbolic metamaterials.

## 1. Introduction

Artificial media can support electromagnetic modes that do not occur in conventional materials making them attractive as building blocks for photonic devices [1]. One such application is thermal engineering where control over the spectral and angular width of thermal radiation along with emissivity can lead to multitude of applications in energy conversion [2], radiative heat transfer [3], thermal stamping [4] and thermal sinks [5]. Simple composite materials offer a possible solution for broadband applications [6] however spectral selectivity and angular control leads us to artificial structural resonances as in photonic crystals [7] or artificial material responses using metamaterials [8–10].

It was conventionally held that the emissivity of a body can be engineered only to its maximum value as given by the black body limit [5]. However, advances in nano-characterization and analysis of surface modes of naturally occurring materials have shown that this maximum value can be exceeded at a resonant frequency [11]. On the other hand, the emissivity can exceed unity in a broad bandwidth using non-resonant artificial media with hyperbolic dispersion [12,13] (indefinite media [14]). For detailed reviews, a history of such hyperbolic metamaterials (h-MMs) and pertinent references, please see [15,16].

A number of recent interesting papers have studied near-field thermal engineering in h-MMs using effective medium theory (EMT) [17–20]. Here we go beyond the EMT and consider nanoscale radiative heat transfer between practical h-MMs. We show that in the mid-infrared range phonon-polaritonic metamaterials using silicon carbide and silicon dioxide multilayers can lead to super-planckian heat transfer between h-MMs. However, for practical applications such as thermophotovltaics high temperature heat transfer in the near-infrared range is required. This arises due to compatibility issues with low bandgap photovoltaic cells for energy conversion and also increased device efficiency at higher temperatures [21]. To this end, we introduce h-MMs based on plasmonic materials with a high melting point. We show that the heat transfer can be super-planckian in the wavelength range between 1 and 3 microns compatible with low bandgap cells for thermophotovoltaic applications. Throughout the paper, we take into account the role of dispersion, absorption and finite unit cell size to validate our effect. This work paves the way for high temperature thermal engineering applications of h-MMs.

## 2. Generalized Kirchhoff's laws

We first consider two homogenous half-space mediums (labeled by 1 and 2) separated by a vacuum gap with width $d$. One medium is at local equilibrium with $T_1$ and the other with $T_2$. Within the framework of Rytov's fluctuational electrodynamics, Polder and Van Hove [22] first derived the general expressions of the heat flux between the two media $H(d,T_1,T_2) = \int_0^\infty (d\omega/2\pi)(\Theta(\omega,T_1) - \Theta(\omega,T_2))S(\omega,d)$ with

$$S(\omega,d) = \sum_{j=s,p} \left\{ \int_0^{k_0} \frac{d^2 k_\parallel}{4\pi^2} \frac{(1-|r_j^{01}|^2)(1-|r_j^{02}|^2)}{|1-r_j^{01}r_j^{02}e^{2ik_z d}|^2} + \int_{k_0}^{\infty} \frac{d^2 k_\parallel}{4\pi^2} e^{-2\mathrm{Im}(k_z)z} \frac{4\,\mathrm{Im}(r_j^{01})\,\mathrm{Im}(r_j^{02})}{|1-r_j^{01}r_j^{02}e^{2ik_z d}|^2} \right\} \quad (1)$$

Here, $\Theta(\omega,T) = \hbar\omega/(\exp(\hbar\omega/k_B T)-1)$ is the mean energy of a harmonic oscillator, $k_0 = \omega/c$ is the free space wave-vector, $k_\parallel = (k_x, k_y)$, $k_\rho = \sqrt{k_x^2 + k_y^2}$, $k_z = \sqrt{k_0^2 - k_\parallel^2}$, $j = s, p$ accounts for the s and p polarizations, $r_j^{0i}$ are the Fresnel reflection coefficients between vacuum (labeled by 0) and the medium ($i=1,2$) for s and p waves. The contribution of propagating waves ($k_\parallel < k_0$) and evanescent waves ($k_\parallel > k_0$) are naturally separated in this expression. The propagating wave term is Kirchhoff's law taking the multiple reflections into account. Here $\mathrm{Im}(r_j^{0i})$ in the evanescent waves term can be interpreted as the generalization of emissivity to the near field [23]. We note $\mathrm{Im}(r_j^{0i})$ are also proportional to the near field local density of states (LDOS) proposed by Pendry [4] and is related to the tunneling and subsequent absorption of energy carried by evanescent waves. Recent work has shown that the heat flux between two planar structures just depends on the scattering matrix, regardless of the inner structure of the half space medium [24]. We utilize this approach for planar multilayer h-MMs. The broadband enhancement in the near field LDOS of h-MMs can be utilized to engineer the heat transfer at the nanoscale [25,26]. We will use both the approximate effective medium theory and the exact Bloch theorem to calculate the reflection coefficients and compare the heat transfer properties of h-MMs.

## 3. Phonon-polaritonic h-MMs

### *3.1 Effective medium theory*

We note that there are a few different designs and material combinations that lead to hyperbolic dispersion [15,16]. The unique properties of this metamaterial stem from its hyperboloidal dispersion relation as opposed to spherical for conventional media. This leads to the existence of high-$k$ modes (large wavevector propagating waves which are normally evanescent in vacuum) which can be understood as collective Bloch plasmon oscillations in practical designs. However, a shift in material choice from conventional plasmonic building blocks like silver and gold is needed for analyzing thermal properties since exciting such bulk plasmonic modes requires high energies (temperatures). One choice is a phonon-polaritonic metamaterial where the high-$k$ modes arise due to coupled phonon-polaritons in the nanostructured medium. We consider here a multilayer combination of silicon dioxide ($SiO_2$) and silicon carbide (SiC) which has a metallic response in the Reststrahlen band due to phonon polaritons. We note that this realization formed the testbed for the first complete characterization of the modes of hyperbolic media due to their low loss [27]. The modes of this h-MMs can be excited at relatively lower temperatures (400-500K) when the peak of black body emission lies within the Reststrahlen band of SiC.

To understand the thermal properties of phonon-polaritonic h-MMs we need to focus only on the Reststrahlen band of SiC where it is metallic. The multilayer structure shows a host of different electromagnetic responses as predicted by EMT $\varepsilon_{xx} = \varepsilon_{yy} = \varepsilon_m \rho + \varepsilon_d (1-\rho)$ and $\varepsilon_{zz} = \varepsilon_m \varepsilon_d / (\varepsilon_m (1-\rho) + \varepsilon_d \rho)$, here $\rho$ is the fill fraction of the metallic medium. In Fig. 1(a), we plot the isofrequency surfaces of this metamaterial achieved at different frequencies. The classification is done using the isofrequency surface of extraordinary waves in the effective uniaxial medium which follow $k_z^2/\varepsilon_{xx} + (k_x^2 + k_y^2)/\varepsilon_{zz} = \omega^2/c^2$ and are hyperboloidal only when $\varepsilon_{xx} \cdot \varepsilon_{zz} < 0$. We can effectively achieve a type I hyperbolic metamaterial with only one negative component in the dielectric tensor ($\varepsilon_{xx} = \varepsilon_{yy} > 0, \varepsilon_{zz} < 0$), type II hyperbolic metamaterial with two negative components ($\varepsilon_{xx} = \varepsilon_{yy} < 0, \varepsilon_{zz} > 0$), effective anisotropic dielectric ($\varepsilon_{xx} = \varepsilon_{yy} > 0, \varepsilon_{zz} > 0$) or effective anisotropic metal ($\varepsilon_{xx} = \varepsilon_{yy} < 0, \varepsilon_{zz} < 0$).

Fig. 1(b) shows the effective medium dielectric constants of the SiC/SiO$_2$ multilayer structure [18]. The regions of hyperbolic behaviour are identified with arrows. Fig. 2(a) shows the heat transfer spectrum between two h-MMs separated by a vacuum gap calculated by Eq. (1). It is seen that only in regions of hyperbolic behavior we see super-planckian thermal energy transfer in agreement with our previous analytical approximation of thermal topological transitions in near-field energy density [18].

### 3.2 Near-field non-locality

We now introduce the concept of near-field non-locality for h-MMs [28,29]. The macroscopic homogenization utilized to define a bulk electromagnetic response is valid when the wavelength of operation exceeds the unit cell size ($\lambda \gg a$). However, even at such wavelengths if one considers incident evanescent waves on the metamaterial, the unit cell microstructure causes significant deviations from EMT. This is an important issue to be considered for quantum and thermal applications where the near-field properties essentially arise from evanescent wave engineering (high-*k* modes). For the multilayer h-MMs, at distances below the unit cell size, the thermal emission is dominated by evanescent waves with lateral wavevectors $k_x \gg (1/a)$. Since this is above the unit-cell cut off of the metamaterial, the high-*k* modes do not contribute to thermal emission at such distances. It is therefore necessary to consider thermal emission from a practical multi-layer structure taking into account the layer thicknesses. This is shown in Fig. 2(a) calculated using Eq. (1). The unit cell size is 20nm, and we consider a semi-infinite multilayer medium using the formalism outlined in [30]. An excellent agreement between the radiative heat spectrum of the multilayer structure and the EMT calculation is observed. The reason for this agreement is our choice of mid-infrared metamaterials where achieving deep sub-wavelength layers is relatively easy as compared to optical metamaterials. Furthermore, phonon-polaritonic metamaterials have lower losses and higher figure of merit compared to their plasmon-polaritonic counterparts.

Along with non-locality due to a finite unit cell size, surface effects also lead to deviations from EMT. To analyze this, we consider the case when the topmost layer is metallic (MD multilayer) and also when it is dielectric (DM multilayer). It is seen that the surface states at the interface of the topmost layer and vacuum plays a significant role in energy transfer when the top layer is metallic [Fig. 2(a)]. This is similar to case of near-field imaging where surface plasmon polaritons at both the object plane and image plane contribute significantly to resolution enhancement; a phenomenon not captured in EMT descriptions [31]. In Fig. 2(b), we show the wavevector resolved heat transfer for a multilayer metamaterial which clearly elucidates the role of high-$k$ metamaterial states in the heat transfer.

## 3.3 Near-field heat transfer spectroscopy

One possible approach to experimentally measure the topological transition present in the thermal spectrum is by using near-field scanning optical spectroscopy [32]. In this case, the surface layer starts influencing the net signal measured by the tip. It is seen that the topological transition due to bulk modes persists in effective medium and multilayer cases [Fig. 2(a)]. However the contribution from the surface-phonon polariton mode of the top layer of silicon carbide plays a major role in the detected signal. Thus the optimum design for detecting the effect using scanning technologies has to utilize a well-designed top layer of $SiO_2$.

## 4. High temperature h-MMs

Note that conventional plasmonic building blocks for metamaterials such as silver and gold cannot be used for high temperature applications due to their low melting point ($T_{m.p.}$≈1000C). Recently, it was proposed that high temperature plasmonic materials [10] such as aluminum zinc oxide and titanium nitride [33] are a promising route for narrowband thermal emission in the near-IR range ($T_{m.p.}$≈2200C). This is a step towards practical high temperature applications and compatibility with low bandgap thermophotovoltaic cells.

In Fig. 3(a), we show the thermal energy transfer between multilayer metamaterials made of aluminum zinc oxide (AZO, plasmonic metal) and titanium dioxide ($TiO_2$) with optical constants taken from [33]. On comparing with the effective medium constants of Fig. 1(c), we see that the thermal emission is above the black body limit in regions of hyperbolic behavior. Furthermore, we note that in the regions where a flat elliptical dispersion is achieved we also see similar behavior due to the large effective index. In addition, the epsilon-near-zero condition gives a peak in the thermal emission which is due to a surface-plasmon-polariton (SPP) resonance (near 3 μm). Note that there is a good match with effective medium theory and the SPP resonance dominates when the surface layer of the multilayer structure is metallic. In Fig. 3(b), we show the comparison of the net heat transfer with size of the gap. As the gap size is decreased, the thermal energy transfer far exceeds the black-body limit. A more optimized design can lead to higher heat transfer values.

## 5. Conclusion

In conclusion we have predicted super-planckian radiative thermal energy transfer in practical multilayer phonon-polaritonic h-MMs. We also introduced a class of high temperature h-MMs for energy transfer in the near-infrared range relevant to thermphotovoltaic applications. We paid particular attention not only to the effective medium approximation but discussed all non-idealities limiting the super-planckian thermal emission from h-MMs. The multilayer design in the paper is easily achievable using existing fabrication technologies and is an ideal platform for experimentally verifying thermal properties of h-MMs such as thermal topological transitions. This work will also lead to a class of high temperature thermal engineering applications of h-MMs.

## Acknowledgments
The authors acknowledge funding from NSERC, Nanobridge and CSEE.

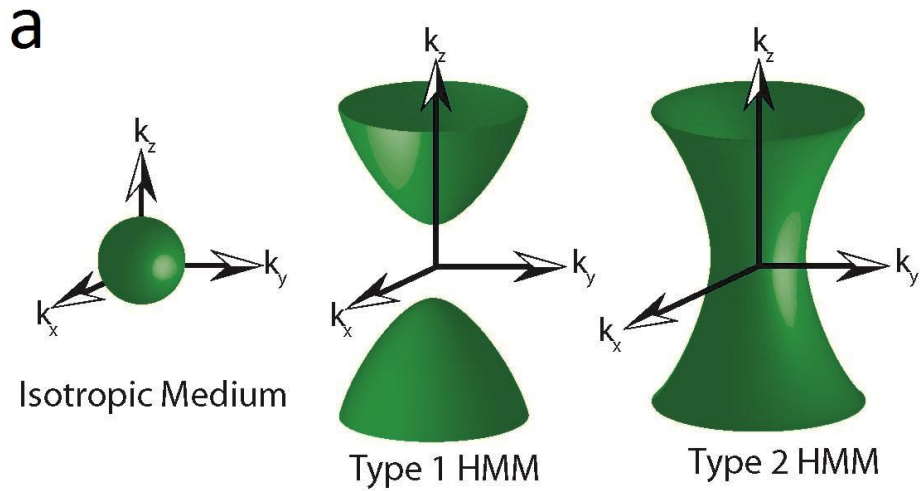

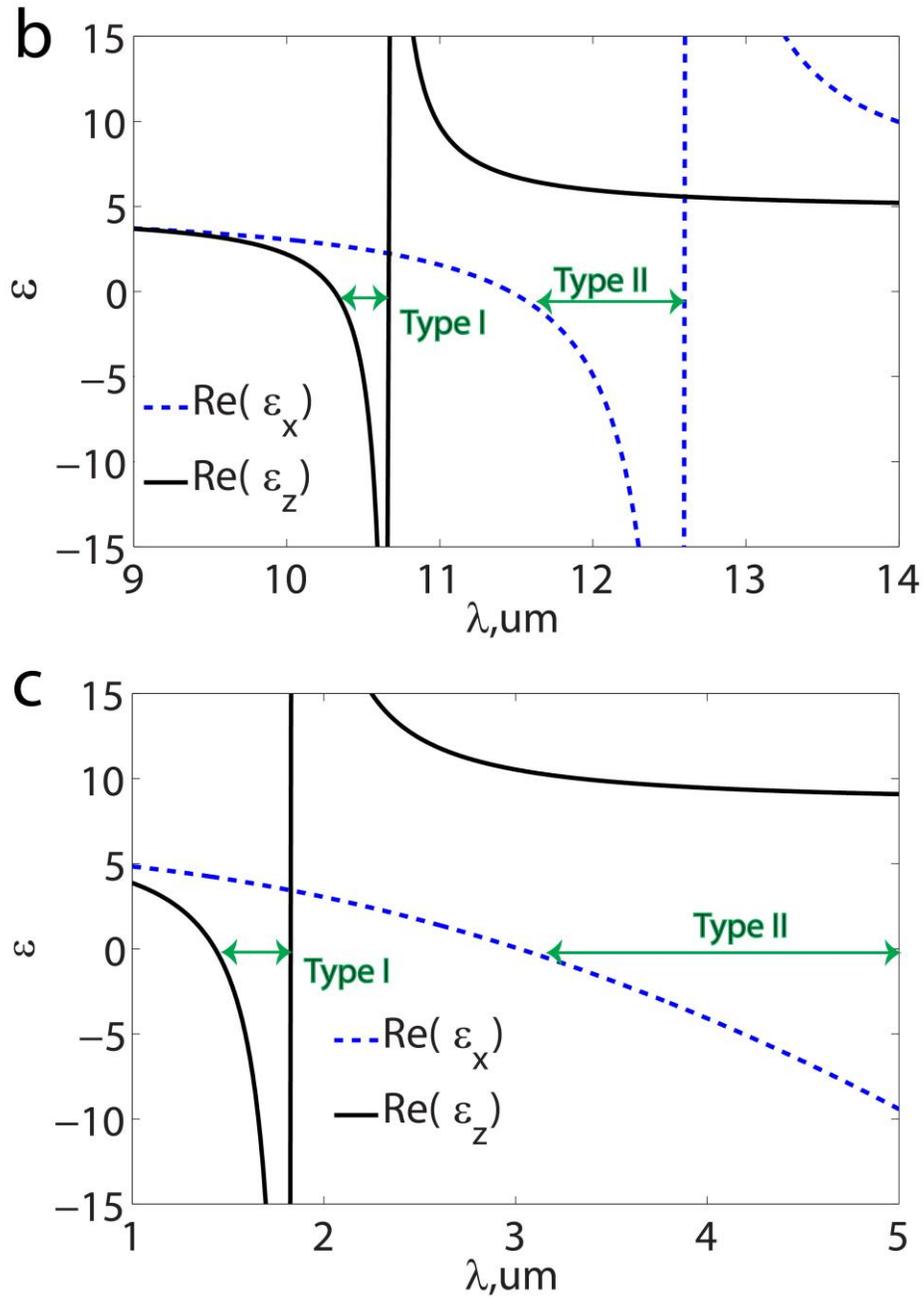

Figure 1: (a) (i) Ellipsoidal isofrequency surface for effective anisotropic dielectric. (ii) type I h-MMs with only one negative component in the dielectric tensor. (iii) type II h-MMs with two negative components. (b) EMT parameters of SiC-SiO$_2$ multilayers with fill fraction=0.3. (c) EMT parameters

of AZO-TiO$_2$ multilayers with fill fraction=0.3. We indicate the hyperbolic regions with arrows in (b) and (c). Note this latter system can be used for high temperature applications.

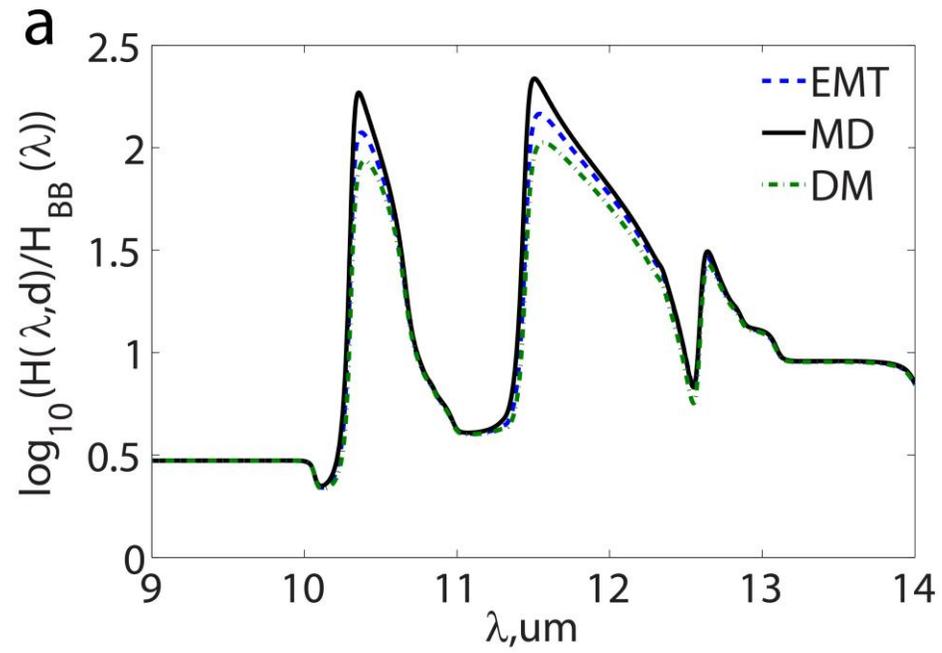

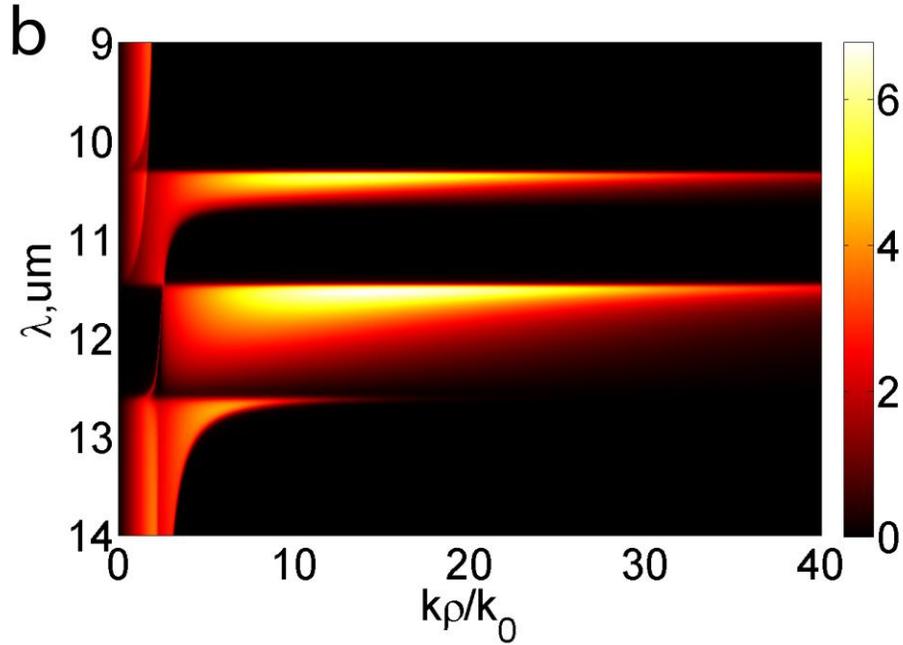

Figure 2: (a) Heat transfer spectrum calculated with EMT, SiC-SiO$_2$ (MD) multilayer, SiO$_2$-SiC (DM) multilayer. The fill fraction of SiC layer is 0.3 and the unit cell size is 20nm. The gap is 100nm. Here we consider two semi-infinite slabs. One slab is at 500K and the other at 0K. There are three distinct peaks. The right peak is due to the pole in the dielectric constant of silicon carbide at the transverse optical phonon frequency. The left and middle one correspond to type I and type II h-MMs, respectively. The higher peaks in the MD curve are due to the surface phonon polaritons of the topmost metallic layers. (b) Wavevector resolved heat transfer of SiC-SiO$_2$ (MD) multilayer normalized to blackbody limit, the three red bright regions clearly show the origin of the three peaks in (a) and the high-k states in hyperbolic regions.

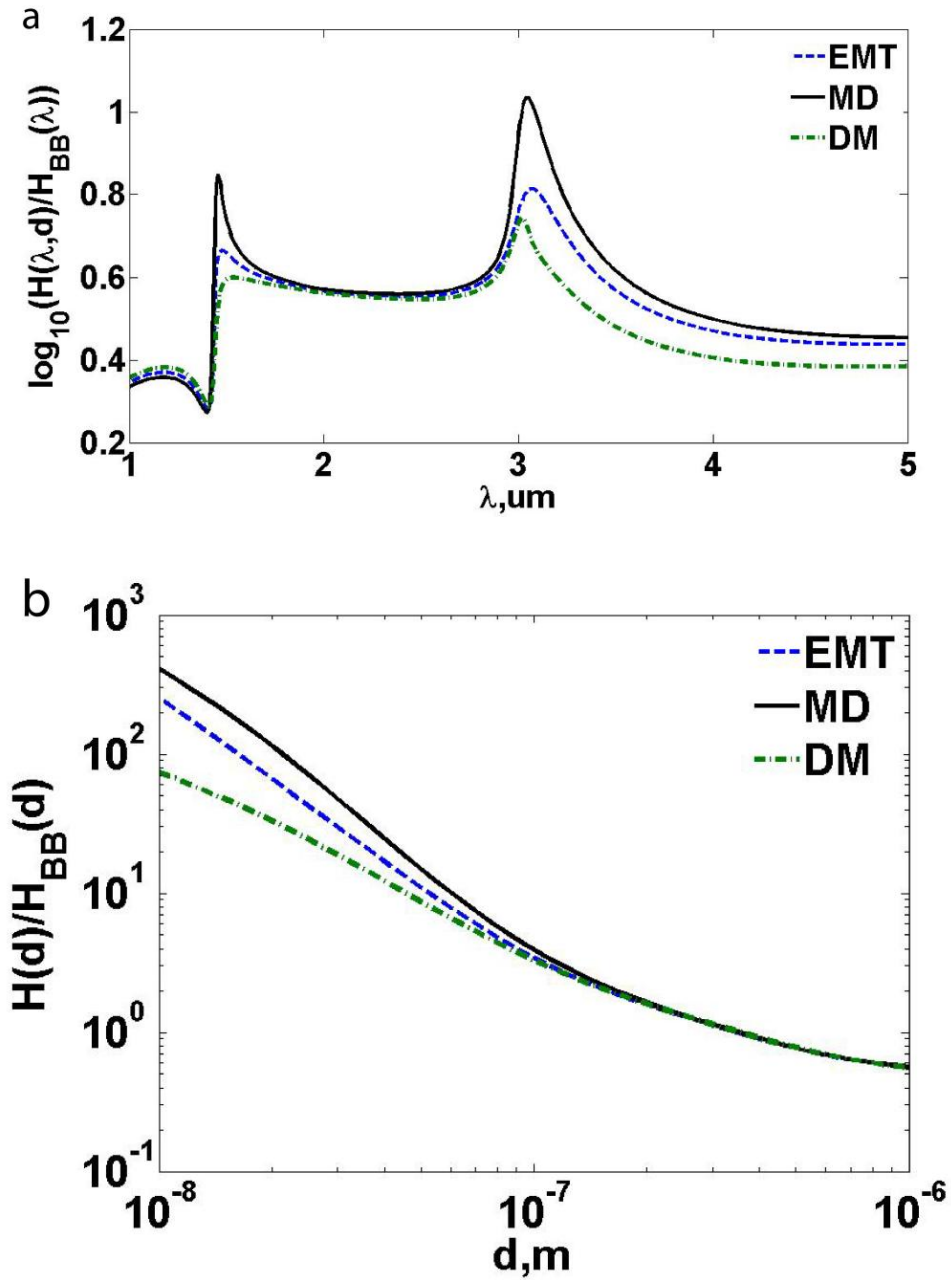

Figure 3: (a) Heat transfer spectrum calculated with EMT, AZO-TiO$_2$ (MD) multilayer, TiO$_2$-AZO (DM) multilayer. The fill fraction of AZO layer is 0.3 and the unit cell size is 20nm. The gap is 100nm. Here we consider two semi-

infinite slabs. One slab is at 1500K and the other at 0K. The left peak is near the epsilon-near-zero transition from elliptical medium to type I hyperbolic region. The right one corresponds to the epsilon-near-zero effect in the type II region at wavelength larger than 3um. The flat region between the two peaks is due to a large effective index in the same wavelength region. The MD curve has higher peaks due to the SPP states of the topmost metallic layers. (b) Net heat transfer as a function of the gap size. At gap sizes much smaller than the operating wavelength (1um), the net heat transfer can be significantly higher than the blackbody limit.